\documentclass[a4paper, superscriptaddress, twocolumn, prl, showpacs]{revtex4-1}

\usepackage{dcolumn}
\usepackage{bm}
\usepackage[colorlinks=true,citecolor=blue,linkcolor=blue,pdfhighlight =/O]{hyperref}
\usepackage{graphicx}
\usepackage{amsmath}
\usepackage{mathtools}
\usepackage{color}
\usepackage{soul}

\newcommand{\mr}[1]{\mathrm{#1}}
\newcommand{\be}{\begin{equation}}
\newcommand{\ee}{\end{equation}}

\newcommand{\figta}{$\left(\mathrm{a}\right)\;$}
\newcommand{\figtb}{$\left(\mathrm{b}\right)\;$}

\newcommand{\figa}{$\left(\mathrm{a}\right)$}
\newcommand{\figb}{$\left(\mathrm{b}\right)$}
\newcommand{\figc}{$\left(\mathrm{c}\right)$}

\newcommand{\kohm}{\;\mr{k}\Omega}

\newcommand{\mk}{\;\mr{mK}}

\newcommand{\mbar}{\;\mr{mbar}}

\newcommand{\ohmcm}{\;\Omega\mr{cm}}

\newcommand{\mum}{\;\mu\mr{m}}

\newcommand{\muev}{\;\mu\mr{eV}}
\newcommand{\nm}{\;\mr{nm}}
\newcommand{\s}{\;\mr{s}}

\newcommand{\rtnis}{R_{\mr{T,NIS}}}
\newcommand{\qdotnis}{\dot{Q}_{\mr{NIS}}}
\newcommand{\inis}{I_{\mr{NIS}}}

\newcommand{\nng}{n_{\mr{g}}}

\newcommand{\tb}{T_{\mr{b}}}
\newcommand{\ts}{T_{\mr{S}}}

\newcommand{\vnis}{V_{\mr{NIS}}}

\newcommand{\nns}{n_{\mr{S}}}

\hypersetup{urlcolor=blue}

\begin{document}
% Use the \preprint command to place your local institutional report number in the upper righthand corner of the title page in preprint mode.
% Use the 'preprintnumbers' class option to override journal defaults  to display numbers if necessary
%\preprint{}
\title{Thermal Conductance of a Single-Electron Transistor}
% repeat the \author .. \affiliation  etc. as needed  \email, \thanks, \homepage, \altaffiliation all apply to the current author. Explanatory text should go in the []'s, actual e-mail address or url should go in the {}'s for \email and \homepage.
% Please use the appropriate macro foreach each type of information \affiliation command applies to all authors since the last
% \affiliation command. The \affiliation command should follow the other information
\author{B. Dutta}
\affiliation{Universit\'e Grenoble Alpes, CNRS, Institut N\' eel, 25 Avenue des Martyrs, 38042 Grenoble, France}
\author{J. T. Peltonen}
\affiliation{Low Temperature Laboratory, Department of Applied Physics, Aalto University School of Science, P.O. Box 13500, 00076 Aalto, Finland}
\author{D. S. Antonenko}
\affiliation{Skolkovo Institute of Science and Technology, Skolkovo, 143026 Moscow, Russia}
\affiliation{L. D. Landau Institute for Theoretical Physics, 142432 Chernogolovka, Russia}
\affiliation{Moscow Institute of Physics and Technology, Moscow, 141700, Russia}
\author{M. Meschke}
\affiliation{Low Temperature Laboratory, Department of Applied Physics, Aalto University School of Science, P.O. Box 13500, 00076 Aalto, Finland}
\author{M. A. Skvortsov}
\affiliation{Skolkovo Institute of Science and Technology, Skolkovo, 143026 Moscow, Russia}
\affiliation{L. D. Landau Institute for Theoretical Physics, 142432 Chernogolovka, Russia}
\affiliation{Moscow Institute of Physics and Technology, Moscow, 141700, Russia}
\author{B. Kubala}
\affiliation{Institute for Complex Quantum Systems and IQST, University of Ulm, 89069 Ulm, Germany}
\author{J. K\" onig}
\affiliation{Theoretische Physik and CENIDE, Universit\" at Duisburg-Essen, 47048 Duisburg, Germany}
\author{C. B. Winkelmann}
\affiliation{Universit\'e Grenoble Alpes, CNRS, Institut N\' eel, 25 Avenue des Martyrs, 38042 Grenoble, France}
\author{H. Courtois}
\affiliation{Universit\'e Grenoble Alpes, CNRS, Institut N\' eel, 25 Avenue des Martyrs, 38042 Grenoble, France}
\author{J. P. Pekola}
\affiliation{Low Temperature Laboratory, Department of Applied Physics, Aalto University School of Science, P.O. Box 13500, 00076 Aalto, Finland}
%\thanks{}
%Collaboration name if desired (requires use of superscriptaddress option in \documentclass). \noaffiliation is required (may also be used with the \author command).
%\collaboration can be followed by \email, \homepage, \thanks as well. \collaboration{}
\date{\today}

\begin{abstract}
We report on combined measurements of heat and charge transport through a single-electron transistor. The device acts as a heat switch actuated by the voltage applied on the gate. The Wiedemann-Franz law for the ratio of heat and charge conductances is found to be systematically violated away from the charge degeneracy points. The observed deviation agrees well with the theoretical expectation. With large temperature drop between the source and drain, the heat current away from degeneracy deviates from the standard quadratic dependence in the two temperatures.
\end{abstract}

%\keywords{}
\maketitle

The flow of heat at the microscopic level is a fundamentally important issue, in particular if it can be converted into free energy via thermoelectric effects \cite{Sothmann2014}. The ability of most conductors to sustain heat flow is linked to the electrical conductance $\sigma$ via the Wiedemann-Franz law: $\kappa/\sigma = L_0T$, where $\kappa$ is the heat conductance, $L_0=\frac{\pi^2k_\text{B}^2}{3e^2}$ the Lorenz number and $T$ the temperature. While the understanding of quantum charge transport in nano-electronic devices has reached a great level of maturity, heat transport experiments are lagging far behind \cite{Dubi2009}, for two essential reasons: (i) unlike charge, heat is not conserved and (ii) there is no simple thermal equivalent to the ammeter. Heat transport can nevertheless give insight to phenomena that charge transport is blind to \cite{Chiatti2006,Banerjee2016} and, remarkably, a series of experiments has demonstrated the very universality of the quantization of heat conductance, regardless of the carriersÕ statistics \cite{Schwab2000,Meschke2006,Molenkamp1992,Chiatti2006,Hoffman2008,Jezouin2013,Banerjee2016, ScienceReddy, Mosso17}.
 
As device dimensions are reduced, electron interactions gain capital importance, leading to Coulomb blockade in mesoscopic devices in which a small island is connected by tunnel junctions. A metallic island connected to a source and a drain through tunnel junctions exceeding the Klitzing resistance $R_\text{K} = h/e^2$ and under the influence of a gate electric field constitutes a Single-Electron Transistor (SET) \cite{Averin}. The charging energy of the island by a single electron writes $E_\text{C} = e^2/2C$ where $C$ is the total capacitance of the island. It defines the temperature and bias thresholds below which single-electron physics appears. In the regime where charge transport is governed by unscreened Coulomb interactions, the question of the associated heat flow has been addressed by several theoretical studies \cite{PRB-Beenakker,Boese2001,PRB-Taring,PRB-Kubala,Tsaousidou2006,Zianni2007,PRL-Kubala,PRB-Rodionov}. The Wiedemann-Franz law is expected to hold in an SET only at the charge degeneracy points in the limit of small transparency, where the effective transport channel is free from interactions, and is violated otherwise.

\begin{figure}[!t]
\includegraphics[width=0.99\columnwidth]{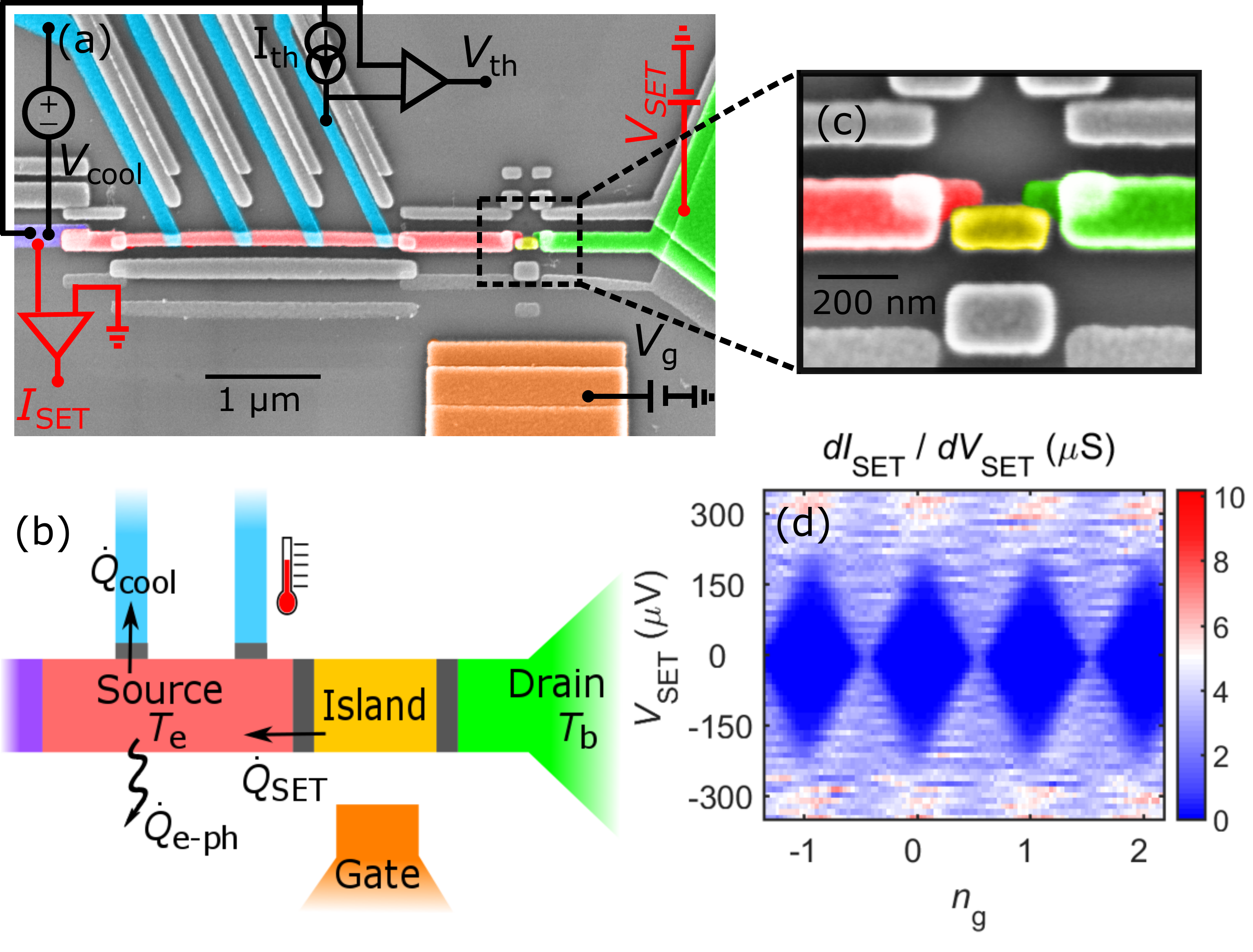}
\caption{(color online). A single-electron transistor and the set-up for the heat transport measurement. (a) False-colored SEM image of the full device. The circuit in red indicates the charge transport set-up, while the black one stands for the heat transport set-up. (b) Schematic of the device, with the different elements shown in colors. (c) Zoomed-in view of the central part of the SET. (d) Differential conductance map of sample A SET at 50 mK against drain-source and gate voltages.}
\end{figure}

In this Letter, we report on the measurements of both the heat and charge conduction through a metallic SET, with both quantities displaying a marked gate modulation. A strong deviation from the Wiedemann-Franz law is observed when the transport through the SET is driven by the Coulomb blockade, as the electrons flowing through the device are then filtered based on their energy.

Figure 1a is a colored scanning electron micrograph of one of the devices that we have investigated while Fig.\ 1b shows a schematics with the same colors for every element. It includes an SET with a drain made of a bulky electrode that is well thermalized to the bath. In contrast, the source of the SET is connected to its lead through a direct Normal metal-Superconductor (NS) contact, which thermally isolates it due to poor thermal conductivity of a superconductor at low temperature. In addition, four superconducting contacts form Superconductor-Insulator-Normal metal (SIN) junctions. As will be discussed below, the latter can be used either as electronic thermometers or coolers/heaters. Samples were fabricated by three-angle evaporation of Cu (30-45 nm), Al (20 nm) and again Cu (30 nm) \cite{SuppMat}. The Al layer was oxidized in order to form tunnel barriers with the second Cu layer. Still, the drain, island and source are in the normal state as the SET tunnel junctions are based on a short Al strip rendered normal by inverse proximity effect via a clean contact to a long normal (Cu) line \cite{koski11}. The SET island was designed with a small volume in order to render the electron-phonon coupling negligible in the island. 

We report here on two investigated devices with identical geometry but different drain-source tunnel resistance $R_\text{N}$ of 164 k$\Omega$ (sample A) and 52 k$\Omega$ (sample B). Figure 1d shows the differential conductance at 50 mK as a function of both the SET bias $V_\text{SET}$ and the average number $n_\text{g} = C_\text{g} V_\text{g}/e$ of electrons induced electrostatically by the gate potential $V_\text{g}$ on the island. Here $C_\text{g}$ is the capacitance between the gate and the island. Coulomb diamonds (in dark blue) are regions of zero current through the SET. Every diamond is centered around an integer value of $n_\text{g}$ and defines a fixed charge state on the island. At zero bias, the charge conductance is thus vanishing, except in the vicinity of the degeneracy points at half-integer values of $n_\text{g}$. At these points, two charge states have the same energy and the conductance (for small barrier transparency) is half the high-temperature value, which is related to the fact that only these two states are involved. From the map, one can estimate a charging energy $E_\text{C}$ of about 155 and 100 $\mu$eV for sample A and B, respectively.

In the present work, our approach is to study the thermal balance in the source when it is heated or cooled. In every thermal measurement, we ensured that no current is flowing through the SET, so that pure heat transport can be considered. The thermal conductance of the SET is inferred from the heat balance in the source, and then compared to the electrical conductance measured in parallel.

We will consider here that the electron population of the source is in quasi-equilibrium at a well-defined (electronic) temperature $T_\text{e}$. This is justified as the mean electron escape time from this element is longer than the estimated electron-electron interaction time \cite{PRL97-Pothier}. By heating or cooling electrons in the source, its electronic temperature $T_\text{e}$ can be different from the temperature of the phonons thermalized at the bath temperature $T_\text{b}$. We achieve electronic thermometry by measuring the voltage drop across a current-biased NIS junction \cite{giazotto06,APL93-Nahum,PRApp-Feshchenko}, the current set-point being chosen to be low enough in the sub-gap regime ($eV < \Delta$, $\Delta$ being the energy gap of the superconductor) to avoid any significant cooling.

\begin{figure}[!t]
\includegraphics[width=0.99\columnwidth]{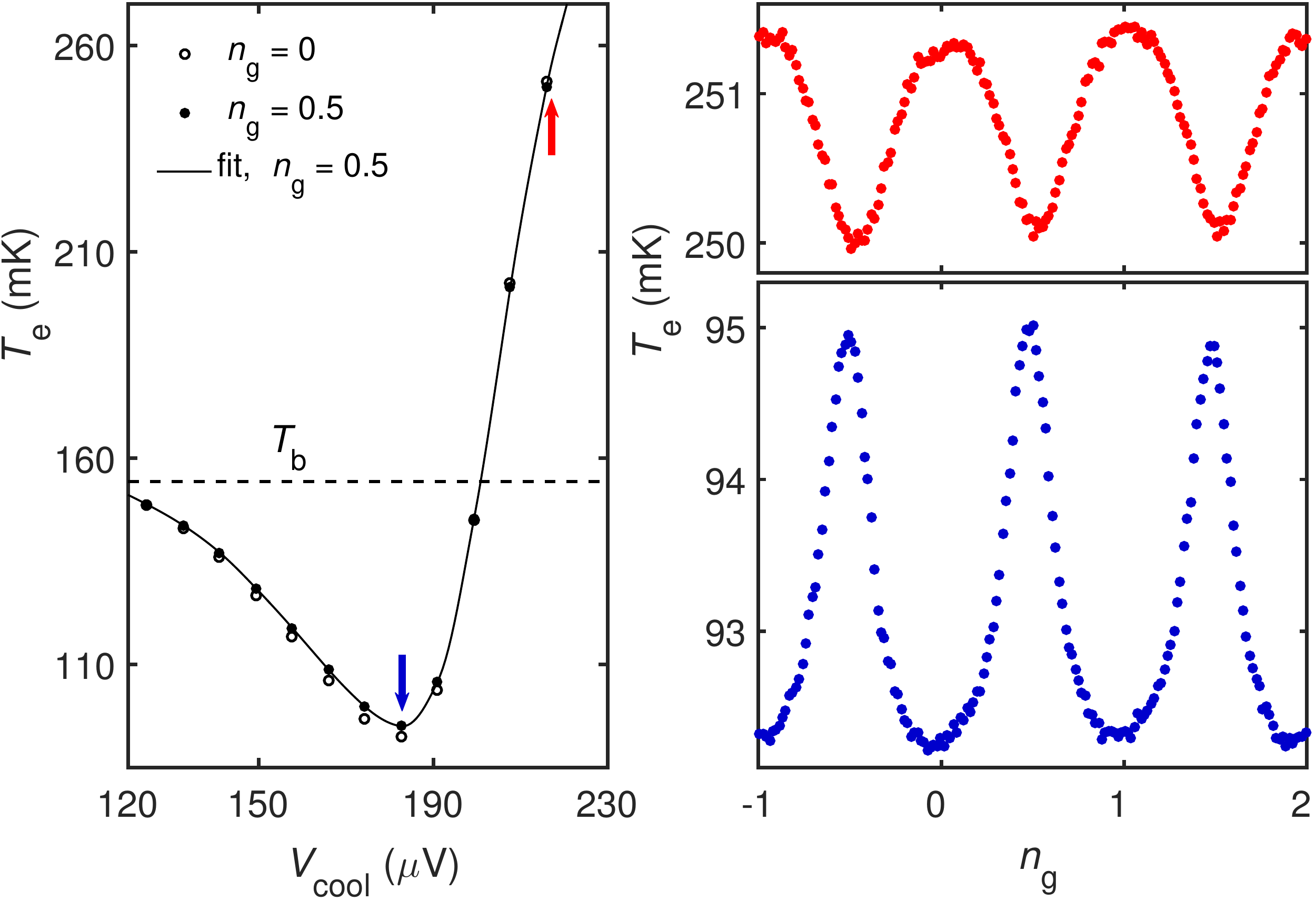}
\caption{Left: Variation of electronic temperature $T_\text{e}$ of sample B source with cooler bias voltage, at gate open ($n_\text{g}$ = 0.5) and gate closed ($n_\text{g}$ = 0) states, at a bath temperature $T_\text{b}$ of 152 mK. The full line is a fit of the gate-open state data, see text. Right: Temperature modulation by the gate voltage expressed in terms of induced charge $n_\text{g}$  in the heating regime (top) and in the cooling regime (bottom) at cooler bias points indicated by the blue and red arrows in the left plot.}
\end{figure}

Indeed, a current bias through a (pair of) NIS junction enables to cool electrons with respect to the phonons \cite{APL95-Nahum,RPP-Muhonen}. This can be understood as a kind of selective evaporation: when the voltage drop is below the energy gap, only higher energy electrons can escape the normal metal. The maximum cooling power is obtained right below the gap in terms of voltage drop across one NIS junction. At a larger voltage, the usual Joule heating is recovered and electrons are heated above the thermal bath temperature.

The cooling and heating of the source electronic bath is illustrated for sample B in Fig.\ 2 left. Here one NIS junction to the source is used for thermometry while a second junction acts as a cooler used for cooling/heating. At a low cooler bias $V_\text{cool}$, the electronic temperature $T_\text{e}$ is below the bath temperature $T_\text{b}$ of 152 mK (indicated by a horizontal dashed line in Fig. 2 left) so that cooling is achieved. The maximum temperature reduction of about 50 mK is reached at a potential drop $V_\text{cool}$ of about 190 $\mu$eV, close to the gap $\Delta$ for Al. A larger cooling is obtained when the gate potential is adjusted so that electron transport through the SET is blocked ($n_\text{g} = 0$) and so is thermal transport through it. At higher bias of the cooler ($V_\text{cool}>\Delta$), an electron overheating is obtained: $T_\text{e} > T_\text{b}$.  Again, the electron temperature change (here an increase) is larger when the SET is blocked. The electron temperature at a fixed cooler bias but as a function of the gate potential is displayed in Fig.\ 2 right. Clear temperature oscillations are obtained, with an opposite sign for the electron cooling and the over-heating regimes. This demonstrates the contribution of the thermal conductance of the SET to heat transport. 

In order to quantify the thermal conductance through the SET, we describe the thermal balance in the source following a thermal model depicted in Fig.\ 1b. In this model, the electron bath in the source receives the power $\dot{Q}_\text{cool}$ from the cooler junction, with a positive or negative sign corresponding to cooling or heating respectively. It can be calculated from \cite{giazotto06} $\dot{Q}_\text{cool}=\frac{1}{e^2R_\text{cool}} \int_{-\infty}^{\infty} (E-eV_\text{cool}) n_\text{S}(E) [f_\text{source}(E-eV_\text{cool})-f_\text{S}(E)]dE-\dot{Q}_0$, where $R_\text{cool}$ is the tunnel junction resistance of the cooler, $n_\text{S}(E)$ is the (BCS) density of states of the superconductor, $f_\text{source,S}(E)$ is the thermal energy distribution function in the source or the S lead of the cooler at respective temperatures $T_\text{e}$ and $T_\text{S}$. The parasitic power $\dot{Q}_0$ takes into account imperfect thermalization of the electrical connections. The main energy relaxation channel for the source electrons is the coupling to phonons, with a power following $\dot{Q}_\text{e-ph}=\Sigma \mathcal{V} (T_\text{e}^5-T_\text{ph}^5)$, where $\Sigma$ is characteristic of the material, $\mathcal{V}$ is the volume, and $T_\text{ph}$ is the phonon temperature here assumed to be equal the bath temperature \cite{giazotto06}. Eventually, the SET transmits a power $\dot{Q}_\text{SET}$ to the source. 

Let us first consider the gate-open position $n_\text{g}$ = 0.5, where the two charge states involved in electron transport have the same electrostatic energy. Electron transport is thus (for small barrier transparency) unaffected by electron interaction and the Wiedemann-Franz law is expected to be valid. The power $\dot{Q}_\text{SET}$ can thus be calculated from the measured differential conductance for charge $dI/dV$ at low bias. We use the thermal balance for the source electrons $\dot{Q}_\text{SET} - \dot{Q}_\text{cool} - \dot{Q}_\text{e-ph} = 0$ to extract the cooling/heating power $\dot{Q}_\text{cool}$. Here the electron-phonon coupling power $\dot{Q}_\text{e-ph}$ is calculated using the actual volume $\mathcal{V}$ and a parameter value: $\Sigma$ = 2.8 nW$\,\mu$m$^3\,$K$^{-5}$, close to the expected value for Cu \cite{arxivVilsanen}. The parasitic power $\dot{Q}_0$ is found to be 0.1 fW in agreement with previous works \cite{Meschke2006}. From the values of $\dot{Q}_\text{cool}$ and taking into account the measured electronic temperature, the imposed cooler voltage, and the Al energy gap, one extracts the superconducting lead temperature as a function the bias $V_\text{cool}$ of the cooler. Values, obtained for this temperature $T_\text{S}$ , up to 450 mK \cite{SuppMat} are in line with expectations in a device where no specific care was put for proper quasiparticle evacuation \cite{NJP-Hung}.

\begin{figure}[!t]
\includegraphics[width=0.99\columnwidth]{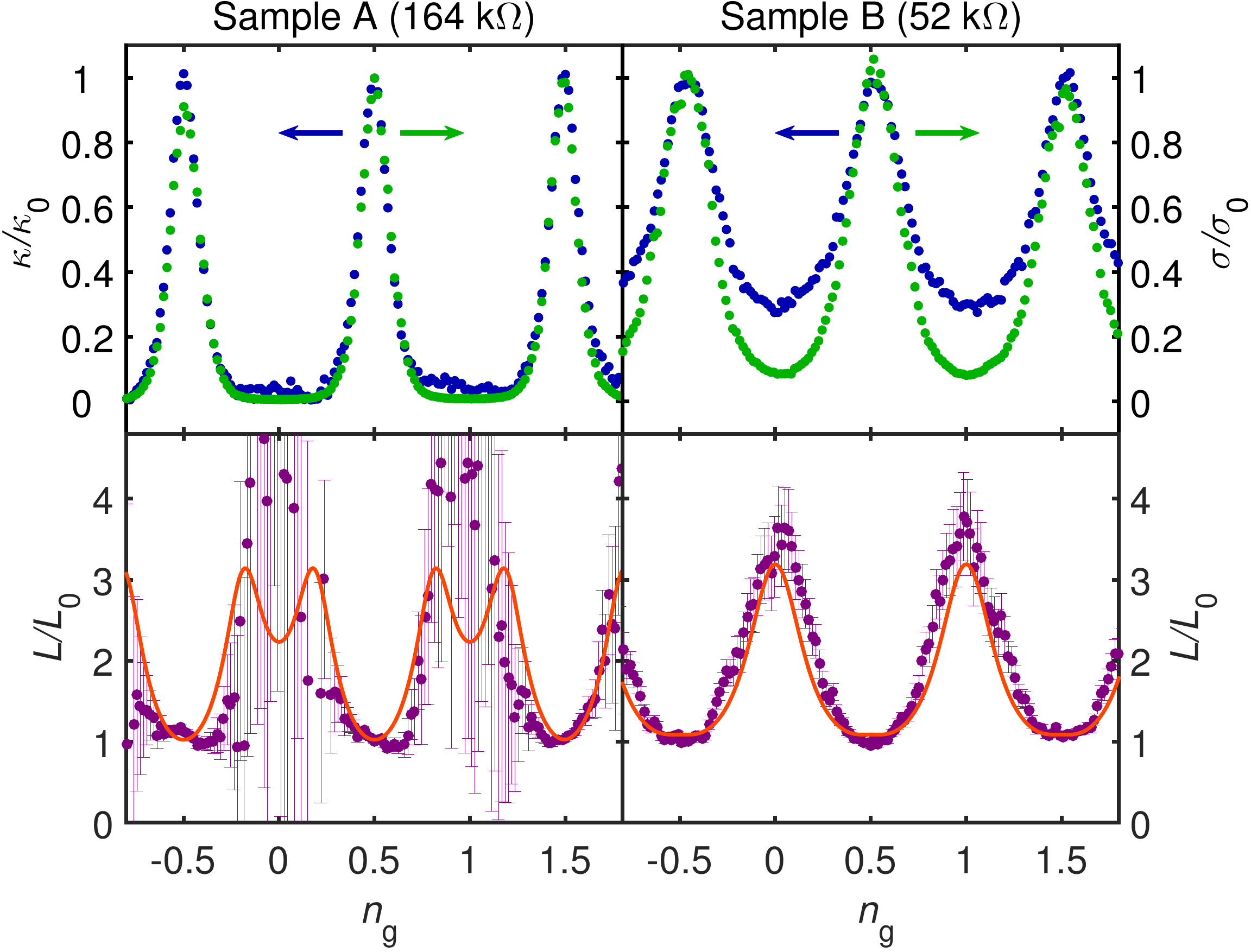}
\caption{Top: Thermal (blue dots) and charge (green dots) conductances of the SET at a bath temperature of 132 mK (left, sample A) and 152 mK (right, sample B) in units of the conductances in the gate open state $\kappa_{0}$ and $\sigma_0$. The thermal flow through the SET was calculated assuming that the Wiedemann-Franz law is fulfilled at gate open. The charge transport was measured at a bias of 22.4 $\mu$V (sample A) and 19.2 $\mu$V (sample B). The heat transport data was acquired by cooling the source electronic bath by 30 mK (sample A) and 22 mK (sample B) below the bath temperature. Bottom: Lorenz ratio (purple dots) defined as $L/L_0$ where $L=\kappa/(\sigma \langle T_\text{m} \rangle)$ for sample A (left) and sample B (right). The error bars are related to the uncertainty in temperature measurement. The Wiedemann-Franz law sets $L=L_0$. The red line is the theoretical prediction based on Ref. \cite{PRL-Kubala}.}
\end{figure}

The preceding analysis at the gate open state provides us with a full knowledge of the thermal behavior of the source, including all physical parameters for electronic cooling and electron-phonon coupling. We now assume that, whatever the gate potential is, the temperature of the superconducting leads of the cooler varies with the cooler's bias as determined above in the gate open case. The measured values of the source electronic temperature $T_\text{e}(n_\text{g})$ are used to calculate the power flowing through the SET as $\dot{Q}_\text{SET} = \dot{Q}_\text{cool} + \dot{Q}_\text{e-ph}$ as a function of $n_\text{g}$. Considering the limit of a small temperature difference, the SET heat conductance is then calculated as $\kappa= \dot{Q}_\text{SET}/(T_\text{b}-T_\text{e})$.

Figure 3a shows both the heat conductance $\kappa$ and the charge conductance $\sigma$ for samples A and B, as a function of the gate potential. Both quantities were measured at the same bath temperature. An SET bias of about 20 $\mu$V and an electron cooling by about 25 mK were used for the charge and the heat transport measurements respectively. The charge conductance is plotted in units of the low-bias gate-open conductance $\sigma_0$. The heat conductance is plotted in units of the Wiedemann-Franz value in the gate-open state $\kappa_{0}=\sigma_0 L_0 \langle T_\text{m} \rangle$. The mean temperature $T_\text{m}=(T_\text{e}+T_\text{b})/2$ is here averaged over the range of induced charge $n_\text{g}$ = 0 to 1. Using $T_\text{m}$ here makes that a linear response is expected in the Wiedemann-Franz regime even for the case of a significant temperature difference $T_\text{e}-T_\text{b}$ \cite{SuppMat}.

For both samples A and B, the charge conductance oscillates with $n_\text{g}$. In the case of sample A (top left), the charge and heat conductances mostly overlap over the full gate potential range. Close to the gate-closed state, the two conductances seem to deviate one from the other but their absolute values are small. In contrast, sample B exhibits a clear deviation from the Wiedemann-Franz law.  At the gate closed state, the heat conductance clearly exceeds the charge conductance multiplied by $L_0T$.

In order to get more insight, let us now consider the Lorenz factor defined as $L/L_0$ with $L=\kappa/(\sigma T_\text{m})$. The Wiedemann-Franz law sets a Lorenz factor equal to unity. In contrast, for sample B the Lorenz factor (Fig.\ 3 bottom right) oscillates between 1 at gate-open state and about 4 at gate closed state. This is the main result of this work. Sample A shows essentially the same behavior over the gate potential range where it can be accurately determined whereas error bars are very large in the vicinity of the gate-closed state due to vanishingly small conductances. We obtained similar results for the whole range of bias points of the cooler, both in the cooling and the heating regimes \cite{SuppMat}.

The physical origin of the violation of the Wiedemann-Franz law resides in the energy selectivity of electron transport through an SET. As a consequence of this, the population of electrons flowing through the SET is non-thermal. For instance, at the gate-closed state, only electrons with an energy (counted from the Fermi level) above the charging energy $E_\text{C}$ contribute to the zero-bias SET conductance. These electrons obviously carry the same (electron) charge but a higher energy. Thus the heat conductance does not decay due to interactions as much as the charge conductance does and the Lorenz number exceeds its basic usual value $L_0$. Electron co-tunneling can counter-balance this, as it involves electrons with an energy close to the Fermi level. The cross-over to the co-tunneling regime shows up at the gate-closed state as a maximum of the Lorenz factor at a temperature $T \approx 0.1 E_\text{C}/k_\text{B}$ \cite{PRL-Kubala}. 

We have calculated the theoretical Lorenz factor for our samples using the theory of Ref. \cite{PRL-Kubala}. Figure 3 bottom shows as full lines the calculated Lorenz factor in parallel with the experimental data. As for parameters, we used the calculated values of $k_\text{B} T/E_\text{C} \approx$ 0.06 and 0.12 for sample A and B respectively and the measured values of the SETs conductance. The theoretical prediction and the experimental data match very well, within error bars. For sample A, the calculated Lorenz number shows a relative minimum in the gate-closed state, which cannot be checked in the experiment due to experimental uncertainties.

Further, we investigated the power $\dot{Q}_\text{SET}$ flowing through the SET beyond the regime of small temperature differences. In the linear regime, the thermal conductance $\kappa$ is proportional to temperature, leading to the quadratic dependence of the heat power on the source ($T_\text{e}$) and drain ($T_\text{b}$) temperatures: $\dot Q_\text{SET} \propto T_\text{b}^2-T_\text{e}^2$. Figure 4 compares experimental data (dots) covering both the cooling and the heating regimes to the latter law, on a log-log plot. In the gate-open case $n_\text{g}$ = 0.5, slope is 1 as assumed in the calibration. Away from the gate-open state, a larger slope is obtained, up to 1.14 at $n_\text{g}$ = 0. Further theoretical work is needed to compare this observation to theoretical predictions.

\begin{figure}[!t]
\includegraphics[width=0.65\columnwidth]{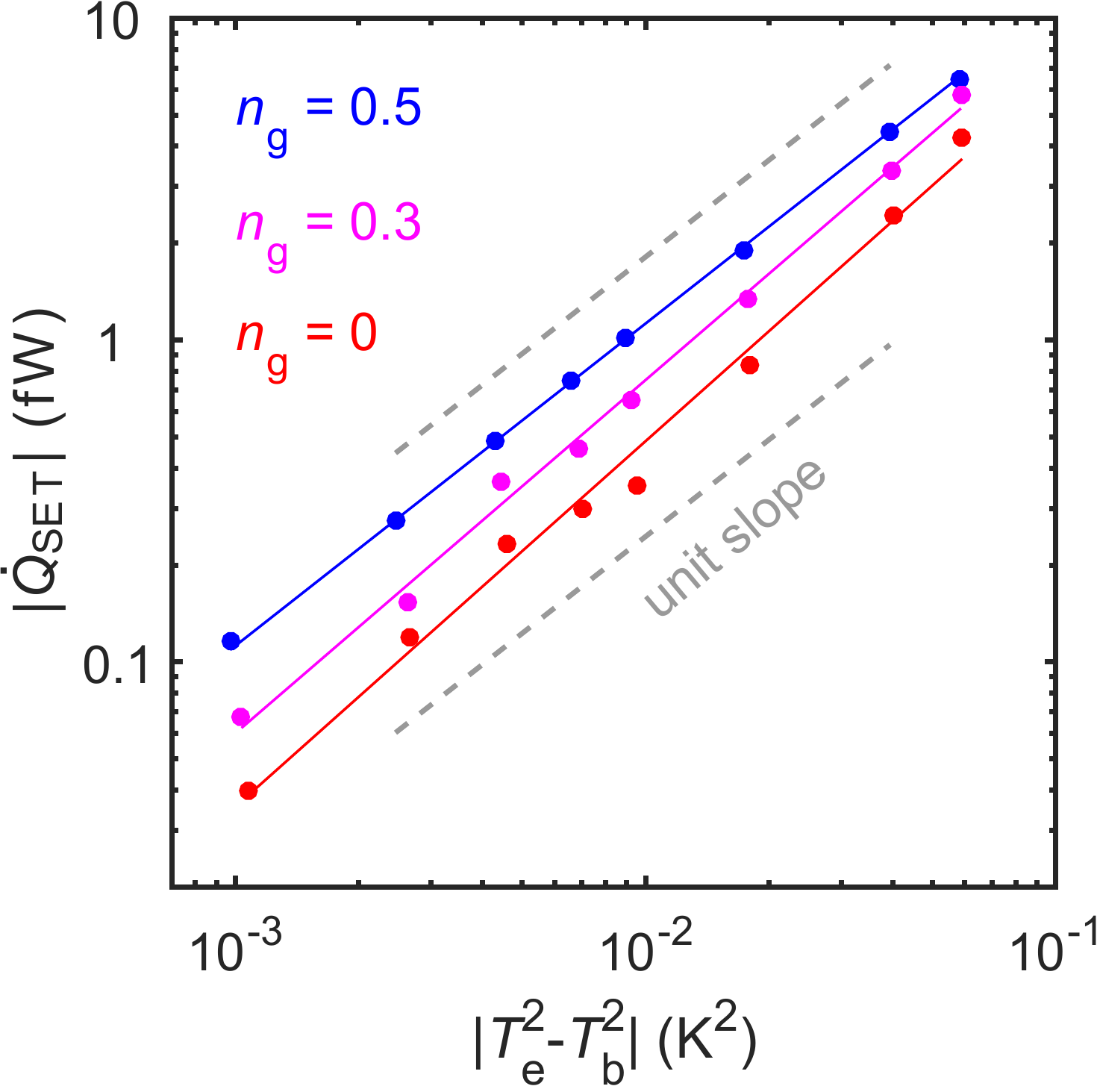}
\caption{Non-linear heat flow through sample B SET at different gate states as a function of the difference of the squared temperatures between the source and the bath (symbols) together with power-law fits (full lines). The slopes are 1.00, 1.10 and 1.14 respectively at gate positions $n_\text{g}$ = 0.5, 0.3 and 0. Unit slope expected for the linear regime of heat transport is shown as a dotted black line.}
\end{figure}

In conclusion, we have demonstrated that the heat transport through an SET can be driven by a gate potential, making the SET a heat switch. The celebrated Wiedemann-Franz law is strongly violated away from the charge degeneracy positions. Our experimental data agrees very well with theoretical predictions. As a prospect, investigating SETs where the island is a quantum dot could exhibit new thermo-electric effects driven by a single energy level \cite{PRL-VanZanten}.

B. D. and J. T. P. contributed equally to this work. We acknowledge financial support from the Nanosciences Fundation, Grenoble. We acknowledge the availability of the facilities and technical support by Otaniemi research infrastructure for Micro and Nanotechnologies (OtaNano). We acknowledge financial support from the Academy of Finland (Projects 272218, 284594 and 275167).

\clearpage
\widetext
\begin{center}
\textbf{\large Supplemental Materials: Thermal Conductance of a Single Electron Transistor}
\end{center}
%%%%%%%%%% Merge with supplemental materials %%%%%%%%%%
%%%%%%%%%% Prefix a "S" to all equations, figures, tables and reset the counter %%%%%%%%%%
\setcounter{equation}{0}
\setcounter{figure}{0}
\setcounter{table}{0}
\setcounter{page}{1}
\makeatletter
\renewcommand{\theequation}{S\arabic{equation}}
\renewcommand{\thefigure}{S\arabic{figure}}
\renewcommand{\bibnumfmt}[1]{[S#1]}
\renewcommand{\citenumfont}[1]{S#1}
%%%%%%%%%% Prefix a "S" to all equations, figures, tables and reset the counter %%%%%%%%%%
\newcommand{\ep}{\epsilon}
\newcommand{\vep}{\varepsilon}
\newcommand{\hc}{{\rm \;h.\,c.\;}}
\newcommand{\sign}{\mathop{\mathrm{sign}}\nolimits}
\renewcommand{\Im}{\mathop{\mathrm{Im}}\nolimits}
\renewcommand{\Re}{\mathop{\mathrm{Re}}\nolimits}
 %\usepackage[style=nature]{biblatex}

%\documentclass[aps,prl,twocolumn,groupedaddress,showpacs]{revtex4-1}

% You should use BibTeX and apsrev.bst for references
% Choosing a journal automatically selects the correct APS BibTeX style file (bst file), so only uncomment the line below if necessary.
%\bibliographystyle{apsrev}
%\usepackage{graphicx}
%\usepackage{float}
%\begin{document}
% Use the \preprint command to place your local institutional report number in the upper righthand corner of the title page in preprint mode.
% Use the 'preprintnumbers' class option to override journal defaults  to display numbers if necessary
%\preprint{}

%\title{Electron Over-Heating Masks Shapiro Steps in Proximity Josephson Junctions}
\title{Thermal Conductance of a Single Electron Transistor}

% repeat the \author .. \affiliation  etc. as needed  \email, \thanks, \homepage, \altaffiliation all apply to the current author. Explanatory text should go in the []'s, actual e-mail address or url should go in the {}'s for \email and \homepage.
% Please use the appropriate macro foreach each type of information \affiliation command applies to all authors since the last
% \affiliation command. The \affiliation command should follow the other information
\author{B. Dutta}
\affiliation{Universit\'e Grenoble Alpes, CNRS, Institut N\' eel, 25 Avenue des Martyrs, 38042 Grenoble, France}
\author{J. T. Peltonen}
\affiliation{Low Temperature Laboratory, Department of Applied Physics, Aalto University School of Science, P.O. Box 13500, 00076 Aalto, Finland}
\author{D. S. Antonenko}
\affiliation{Skolkovo Institute of Science and Technology, Skolkovo, 143026 Moscow, Russia}
\affiliation{L. D. Landau Institute for Theoretical Physics, 142432 Chernogolovka, Russia}
\affiliation{Moscow Institute of Physics and Technology, Moscow, 141700, Russia}
\author{M. Meschke}
\affiliation{Low Temperature Laboratory, Department of Applied Physics, Aalto University School of Science, P.O. Box 13500, 00076 Aalto, Finland}
\author{M. A. Skvortsov}
\affiliation{Skolkovo Institute of Science and Technology, Skolkovo, 143026 Moscow, Russia}
\affiliation{L. D. Landau Institute for Theoretical Physics, 142432 Chernogolovka, Russia}
\affiliation{Moscow Institute of Physics and Technology, Moscow, 141700, Russia}
\author{B. Kubala}
\affiliation{Institute for Complex Quantum Systems and IQST, University of Ulm, 89069 Ulm, Germany}
\author{J. K\" onig}
\affiliation{Theoretische Physik and CENIDE, Universit\" at Duisburg-Essen, 47048 Duisburg, Germany}
\author{C. B. Winkelmann}
\affiliation{Universit\'e Grenoble Alpes, CNRS, Institut N\' eel, 25 Avenue des Martyrs, 38042 Grenoble, France}
\author{H. Courtois}
\affiliation{Universit\'e Grenoble Alpes, CNRS, Institut N\' eel, 25 Avenue des Martyrs, 38042 Grenoble, France}
\author{J. P. Pekola}
\affiliation{Low Temperature Laboratory, Department of Applied Physics, Aalto University School of Science, P.O. Box 13500, 00076 Aalto, Finland}
%Collaboration name if desired (requires use of superscriptaddress option in \documentclass). \noaffiliation is required (may also be used with the \author command).
%\collaboration can be followed by \email, \homepage, \thanks as well. \collaboration{}

\date{\today}
\maketitle
In this supplemental material part, we discuss details of the sample fabrication process, estimation of the SET and NIS probe parameters from independent electrical measurements, as well as the thermal balance. We add complementary experimental data that supports the analysis made in the main paper.

\section{Sample fabrication}

The SET samples are fabricated using a process closely related to the one described in Ref.~\onlinecite{peltonen15}, relying on two rounds of electron beam lithography (EBL) and subsequent metal depositions. The substrate is a p-doped (resistivity $1-30\ohmcm$), single-side polished 4" Si $\langle100\rangle$ wafer with $300\nm$ thermal oxide grown on both sides.

In the first lithography round, a large-area continuous ground plane electrode is patterned and subsequently metallized in an electron beam evaporator as a stack of Ti ($2\nm$) / Au ($30\nm$) / Ti ($2\nm$). After liftoff, the full wafer is coated with an approximately $50\nm$ thick insulating layer of $\mr{Al}_2\mr{O}_3$, grown by atomic layer deposition (ALD). The ground plane electrode starts approximately $20\mum$ away from the SET junctions and the NIS probes [not visible in Fig.\ 1~\figb]. Overlap of the SET and NIS probe leads, defined in the second lithography step, with this ground plane electrode forms an efficient on-chip filter against residual microwave-frequency noise~\cite{pekola10}. The two thin Ti layers help with adhesion to the $\mr{SiO}_2$ substrate and the initial growth of the $\mr{Al}_2\mr{O}_3$ dielectric.

Following the ALD growth, a suspended Ge-based hard mask is prepared for the main EBL step where all the structures shown in Fig.~1~\figtb are defined. The total thickness of the mask is typically $400-500\nm$, whereas the e-beam deposited Ge layer is only $22\nm$ thick, making it possible to optimize the lithography and development steps for reliable formation of the small tunnel junctions. The sacrificial layer under Ge is formed by a spin-coated layer of P(MMA--MAA) copolymer. Immediately after the multi-step dry development process by reactive ion etching (RIE) in $\mr{CF}_4$ and $\mr{O}_2$ plasmas, the sample is loaded into an evaporator equipped with a tiltable sample holder. This allows fabricating both the normal-metal SET and the NIS probes using the same mask and in a single vacuum cycle. The SET is realized using an adaptation of the technique introduced in Ref.~\onlinecite{supp-koski11} as discussed in the following.

\begin{figure}[htb]
\includegraphics[width=0.4\columnwidth]{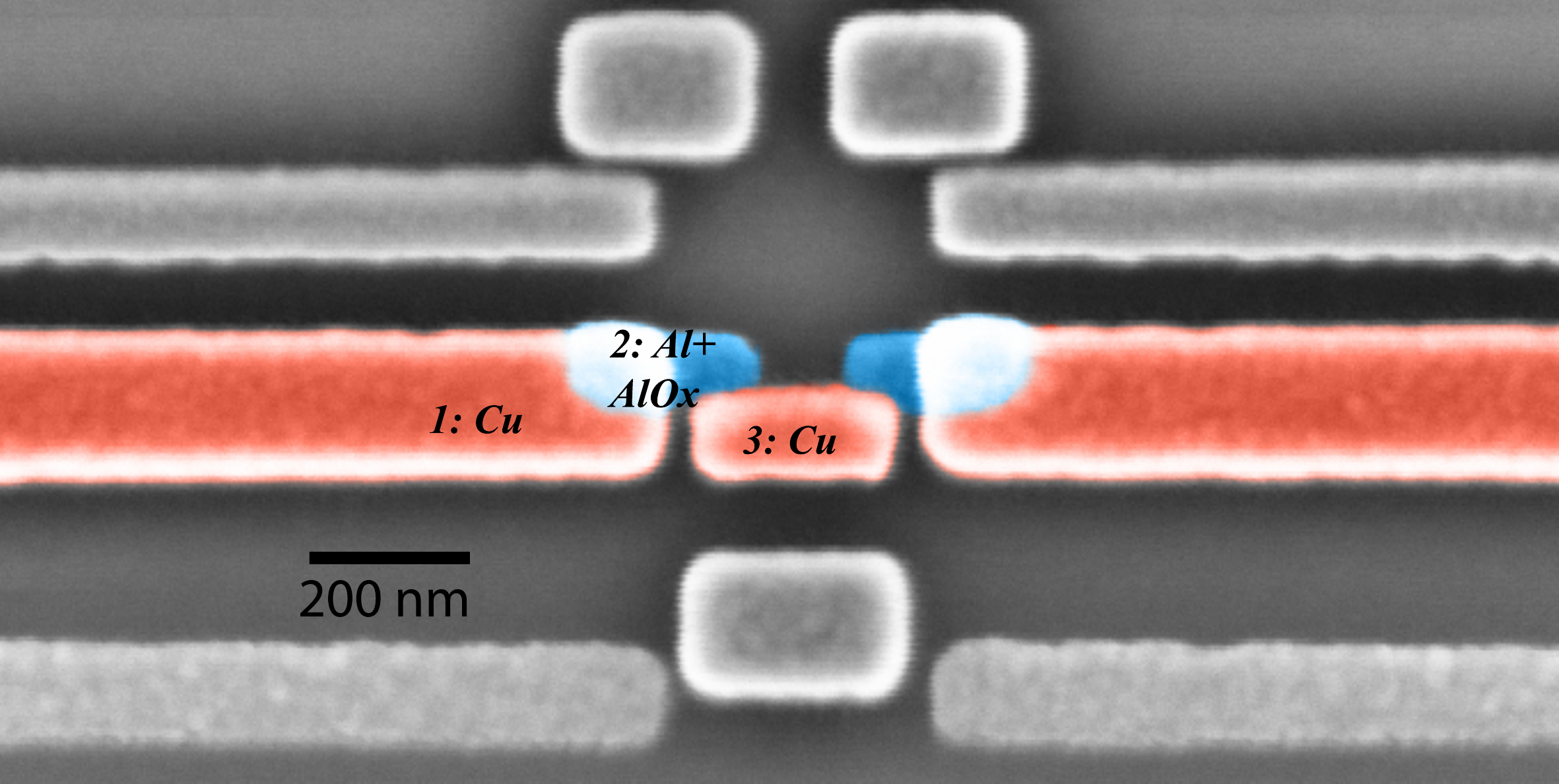}
\caption{False-color scanning electron micrograph of a normal-metallic SET realized with shadow-evaporated Al-proximity junctions. The coloring indicates the materials, and the labels show the order in which the three films are deposited.} \label{fig:proxset}
\end{figure}

First, a $30\nm$ ($45\nm$) thick film of Cu is deposited for sample A (B) with the sample holder set to normal incidence with respect to the evaporation source. As indicated in Fig.~1a, this initial Cu layer forms the SET drain lead as well as the main part of the source electrode. This is immediately followed by the evaporation of a $20\nm$ layer of Al. For this deposition, the sample holder is now tilted to an angle close to $38^{\circ}$, resulting in an effectively close to $15\nm$ thick, downwards [in the orientation of Figs.~1\figtb and~\ref{fig:proxset}] shifted Al copy of the mask pattern. The Al layer forms the two, sub-$200\nm$ long dots, connected to the source and drain Cu regions with transparent metal-to-metal contacts. As observed in Ref.~\onlinecite{supp-koski11}, the Al pieces, with length comparable to the superconducting coherence length, are driven normal due to strong inverse proximity effect from the Cu electrodes. As evident from Fig.~1~\figb, the same Al deposition also forms the S electrodes of the NIS thermometer and cooler/heater [colored light blue in Figs. 1~\figta and~\figc]. To form the $\mr{AlO}_{\mr{x}}$ tunnel barriers for the SET and NIS probe tunnel junctions, the Al layer is subjected to \emph{in-situ} static oxidation immediately after the deposition is completed. This is accomplished by letting typically $1-2\mbar$ of oxygen into the deposition chamber for the duration of $60-90\s$.

To complete the fabrication, a second $30\nm$ layer of Cu is evaporated with the sample now tilted $25^{\circ}$ in the opposite direction compared to the preceding Al deposition. This upwards-shifted copy of the mask pattern forms the SET island [yellow in Figs.\ 1~\figa--\figb] as well as the N electrode of the NIS probes. As a result of the three-angle evaporation through the same mask, three projections of the complete mask pattern will be formed on the substrate. The irrelevant, partially overlapping shadow copies of the various structures, evident in Figs.\ 1 \figtb and~\figc, are shown uncolored in gray.

\section{Sample design}

The length of the narrow and fully superconducting source electrode, i.e., the region where it is not overlapped by the Cu structures and is therefore completely unaffected by the inverse proximity effect, is approximately $3\mum$. At electronic temperatures up to about $250\mk$, this is sufficient for very good thermal isolation of the SET source close to the tunnel junctions \cite{peltonen10}.

The source is made of three Cu elements in series, made of two distinct layers, see Fig.\ 1(a). The Cu--Cu contacts between these elements have low electrical and thermal resistance, as the first Cu film is not significantly oxidized during the formation of the \emph{in-situ} $\mr{AlO}_{\mr{x}}$ tunnel barriers. Therefore, no thermal gradient will develop between the three Cu islands forming the normal-conducting part of the SET source electrode.

We note that in the present work the deposition order for the SET junctions differs from Ref.~\onlinecite{supp-koski11}. Instead of depositing the Al first, followed immediately by Cu in direct contact, and only then oxidizing the Al, we now start with Cu as detailed above. This reversed order for forming the transparent Al-Cu contact allows better experimental control over the tunnel junction transparencies.

\section{SET characterization}

To model the thermal transport properties of the SET, we need to estimate the charging energy $E_\text{C} $ and the tunneling resistance $R_{\mr{T}}$. Assuming symmetric junctions with identical resistances $R_{\mr{T}}$, a straightforward and reliable way to obtain them is to make a fit to the measured minimum and maximum current $I_{\text{SET}}$ at each bias voltage $V_{\text{SET}}$. An example of such calculated envelope curves at $\tb\approx72\mk$, corresponding to $I_{\text{SET}}$ at $\nng=0$ and at $\nng=1/2$, is included in Fig.~S2~for sample B. From this procedure, performed at several bath temperatures $\tb$, we estimate $E_\text{C} \approx 155\muev$ and $R_\text{T} \approx 82\kohm$ for sample A, $E_\text{C} \approx 100\muev$ and $R_{\mr{T}} \approx 26\kohm$ for sample B. For an improved estimate of $E_\text{C}$, we include self-heating at finite currents $I_{\text{SET}}$ and solve the SET source temperature consistently from the thermal balance equation. 

\begin{figure}[htb]
\includegraphics[width=0.4\columnwidth]{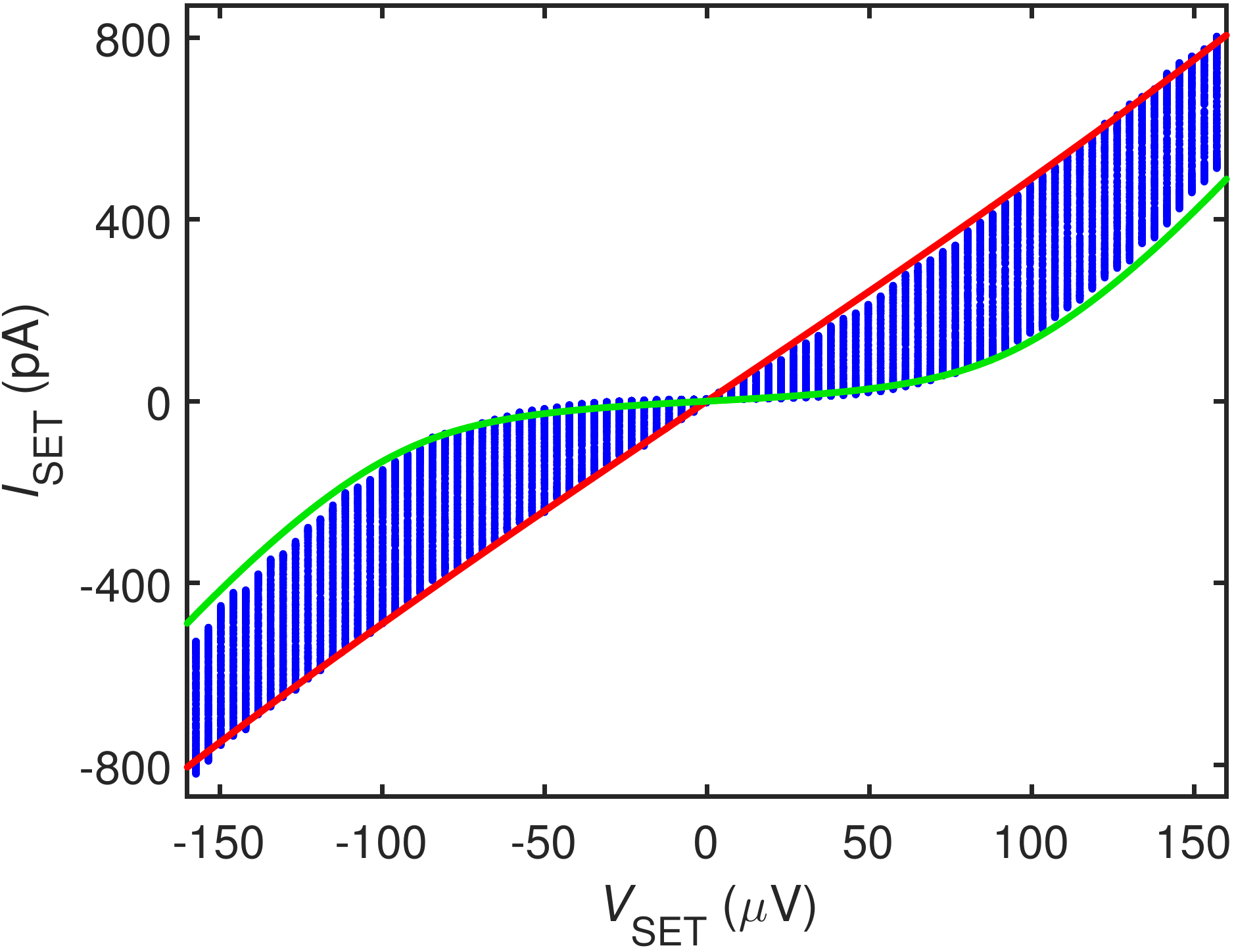}
\caption{
%\figta 
Sample B SET current plotted for different values of the induced charge $n_\text{g}$ as a function of $V_{\text{SET}}$, together with envelope curves calculated at $\nng=0$ (green) and at $\nng=1/2$ (red).
%\figtb Bias- and gate-dependent elevation of SET source temperature $T_1$ used in the calculation of the currents in panel~\figa. The deviations in the gate closed state at $|\vset|\gtrsim 0.3\mv$ are likely due to inaccuracy of the thermal balance, and the assumption of two identical junctions in the SET.
} \label{fig:envelope}
\end{figure}

\section{NIS thermometer and cooler characterization}

Here we show how the main parameters that describe the NIS junction cooler current and its cooling power $\qdotnis$ were estimated. These include the normal state tunnel resistance $\rtnis$, low-temperature superconducting energy gap $\Delta$, and the dimensionless Dynes  broadening parameter $\gamma$. To this end, we fit the measured NIS junction IV characteristic using~\cite{Rajauria07}:
\be
I_\text{NIS}=\frac{1}{2e\rtnis}\int_{-\infty}^{\infty}dE\nns(E)\left[f_\text{source}(E-eV_\text{NIS})-f_\text{source}(E+eV_\text{NIS})\right],\label{inis2}
\ee
showing explicitly that $I_\text{NIS}(V_\text{NIS})=-I_\text{NIS}(-V_\text{NIS})$, and that $I_\text{NIS}$ depends directly only on the electronic temperature $T_\text{e}$ of the SET source electrode. We include the $V_\text{NIS}$-dependence of $T_\text{e}$ via a basic thermal balance.

The low-temperature IV characteristic of the NIS cooler junction of sample B is included in Fig.~S3, both on linear and logarithmic scale, together with the calculated $\inis$. For this sample we obtain $\rtnis\approx 13.2 \kohm$, $\Delta \approx 208\muev$, and $\gamma\approx 8 \times 10^{-4}$. When compared to effects caused by the overheating of the superconducting electrode, the exact value of $\gamma$ or other subgap features of the I--V curve do not play a significant role in modeling the cooling power of the NIS junction at voltages $\vnis$ close to $\Delta/e$.

\begin{figure}[htb]
\includegraphics[width=0.4\columnwidth]{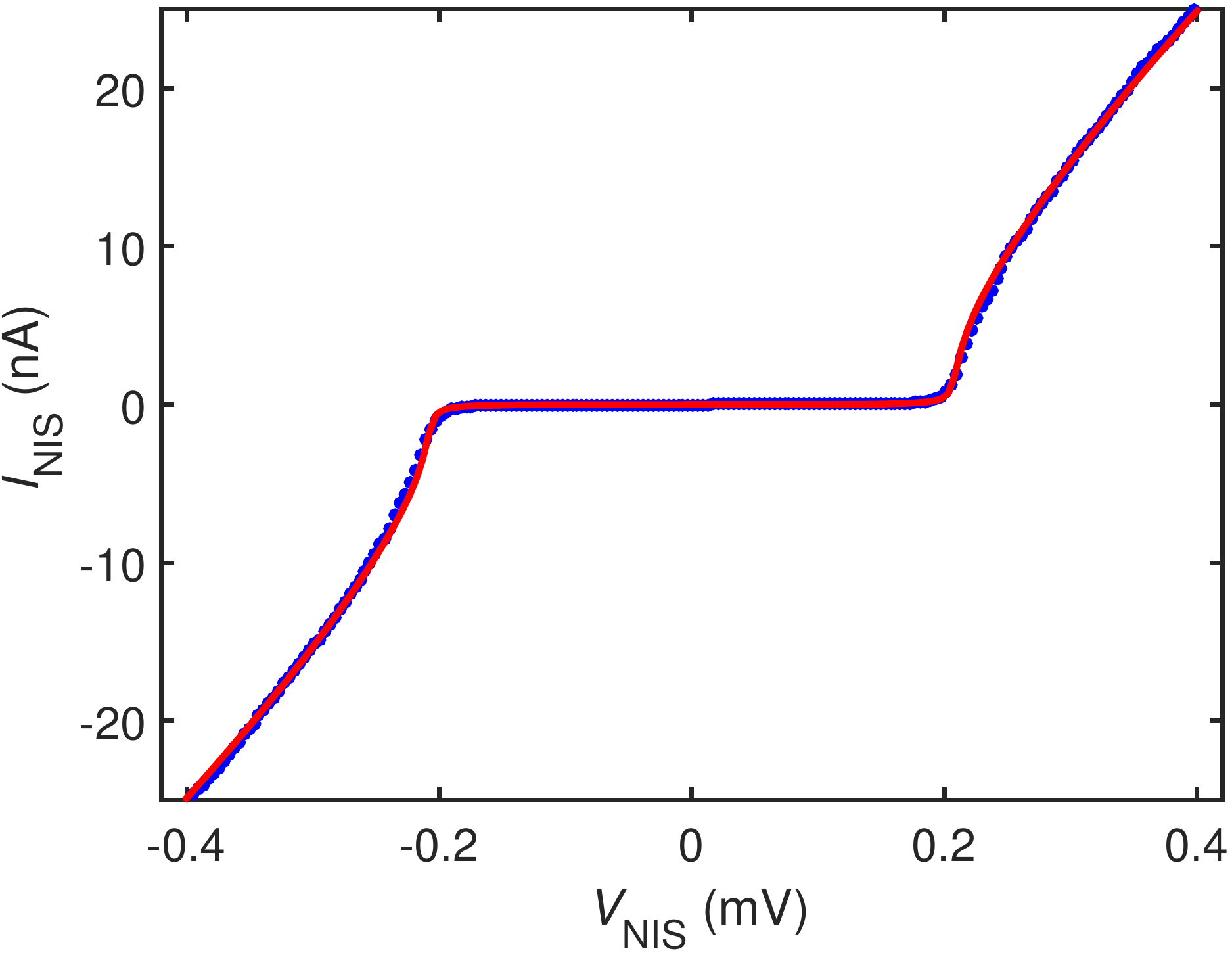}
\includegraphics[width=0.4\columnwidth]{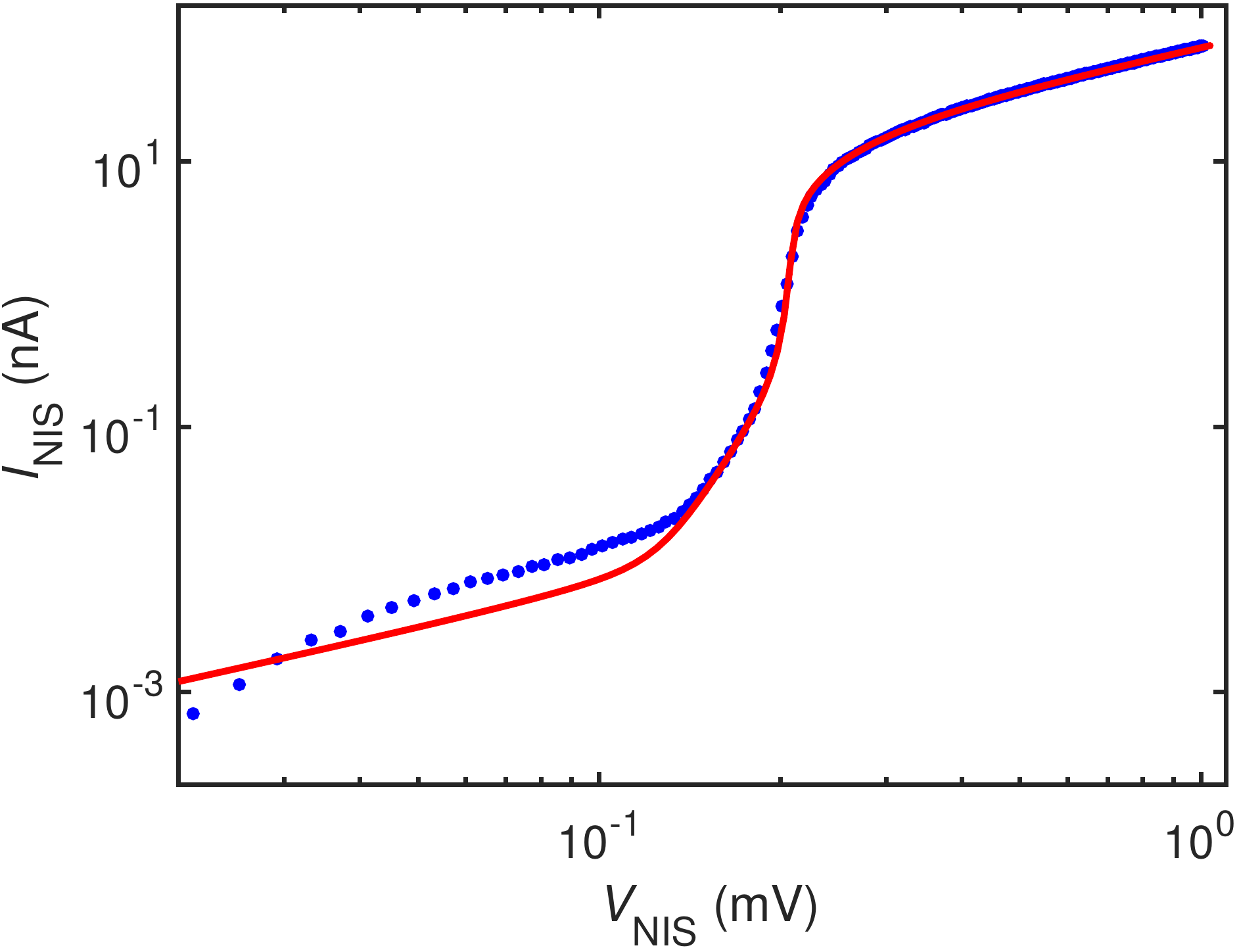}
\caption{Current--voltage characteristic of one NIS junction of sample B on \figta linear and \figtb logarithmic scale. Fits are shown as full red lines.} \label{fig:nisiv}
\end{figure}

\section{Heat balance}

As a simplification for SETs with high normal state resistance, we assume that $\dot{Q}_\text{SET}$ fulfills the Wiedemann-Franz law at $\nng=0.5$ (``gate open''). We then model the actual cooling power of the NIS junction by using an elevated $\ts>\tb$, caused by the injection of non-equilibrium quasiparticles at $e V_\text{cool} \approx \Delta$. At $\tb\approx150\mk$, the superconductor temperature shows values $\ts\approx (200-400)\mk$, cf. Fig.~S4. The order of magnitude of the $\ts$ appears realistic when compared to experiments in similar structures \cite{PRB-Rajauria}. Such a consistent behavior is obtained for bath temperatures up to 300 mK where thermal leakage through the superconducting lead of the source starts to contribute significantly. 

\begin{figure}[htb]
\includegraphics[width=0.4\columnwidth]{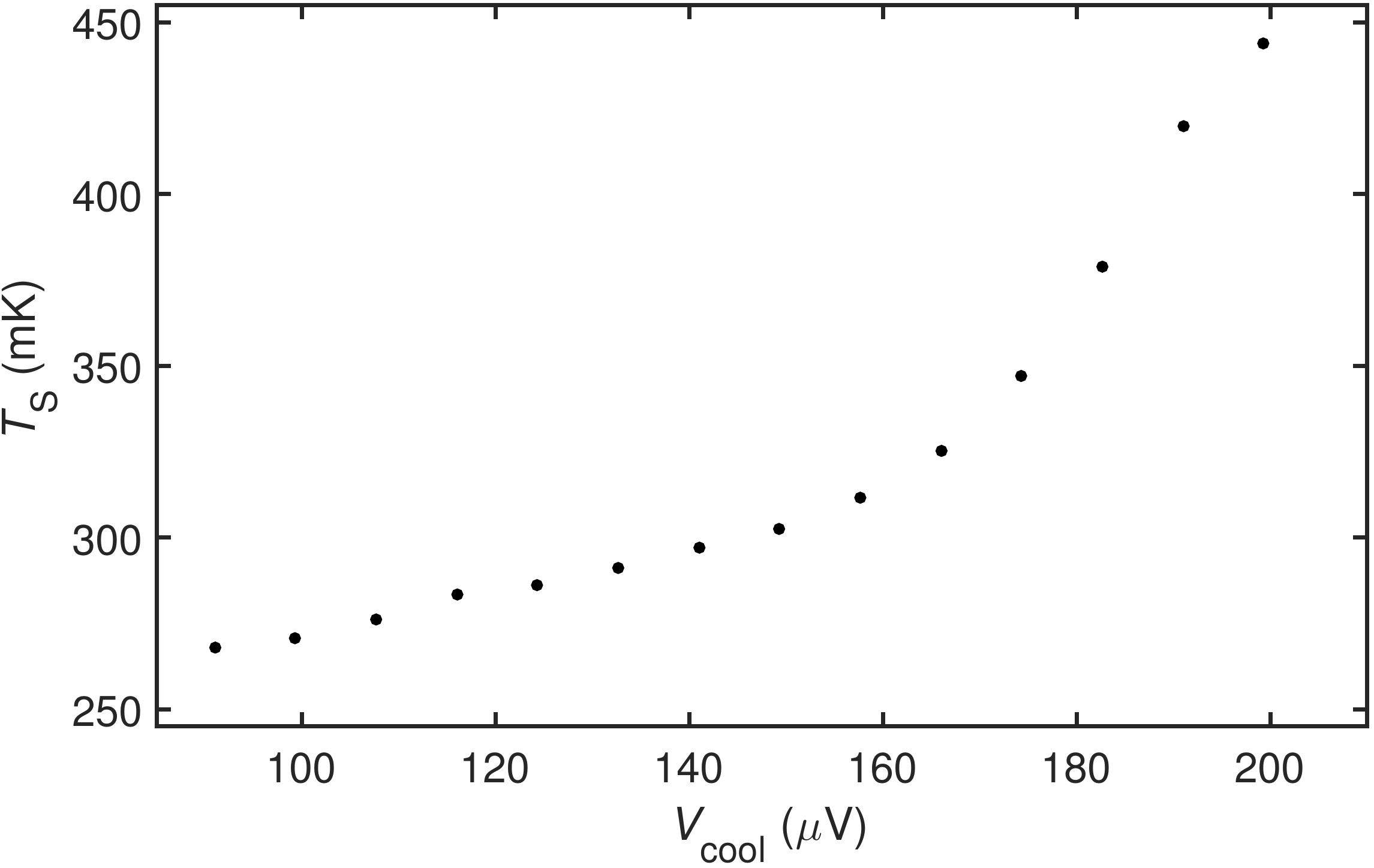}
\caption{Calculated value of the NIS cooler superconductor temperature $\ts$ for sample B, used in the fit of Fig.\ 2 (main paper) data.}
\end{figure}

Once the elevated $\ts$ has been extracted in the above manner, we can use it as well as the \emph{measured} $T_\text{e}$, to extract $\dot{Q}_\text{SET}$ from the heat balance. Notably, this procedure is independent of the model for the SET heat flows -- it rests only on the assumption that $\dot{Q}_\text{SET}(\nng=0.5)$ fulfills the Wiedemann-Franz law.

In the linear regime, the thermal conductance is proportional to temperature: $\kappa=\sigma_0LT$. If one considers a temperature drop between $T_\text{e}$ and $T_\text{b}$, one obtains:
\be
P=\frac{\sigma_0L}{2} (T_\text{b}^2-T_\text{e}^2)=\sigma_0L \frac{T_\text{b}+T_\text{e}}{2}(T_\text{b}-T_\text{e})=\sigma_0L T_\text{m}(T_\text{b}-T_\text{e})
\ee
where $T_\text{m}=(T_\text{b}+T_\text{e})/2$. Considering the Wiedemann-Franz value for the thermal conductance at the average temperature $T_\text{m}$ then enables one to obtain a linear behavior.

\section{Additional experimental material}

\begin{figure}[htb]
\includegraphics[width=0.5\columnwidth]{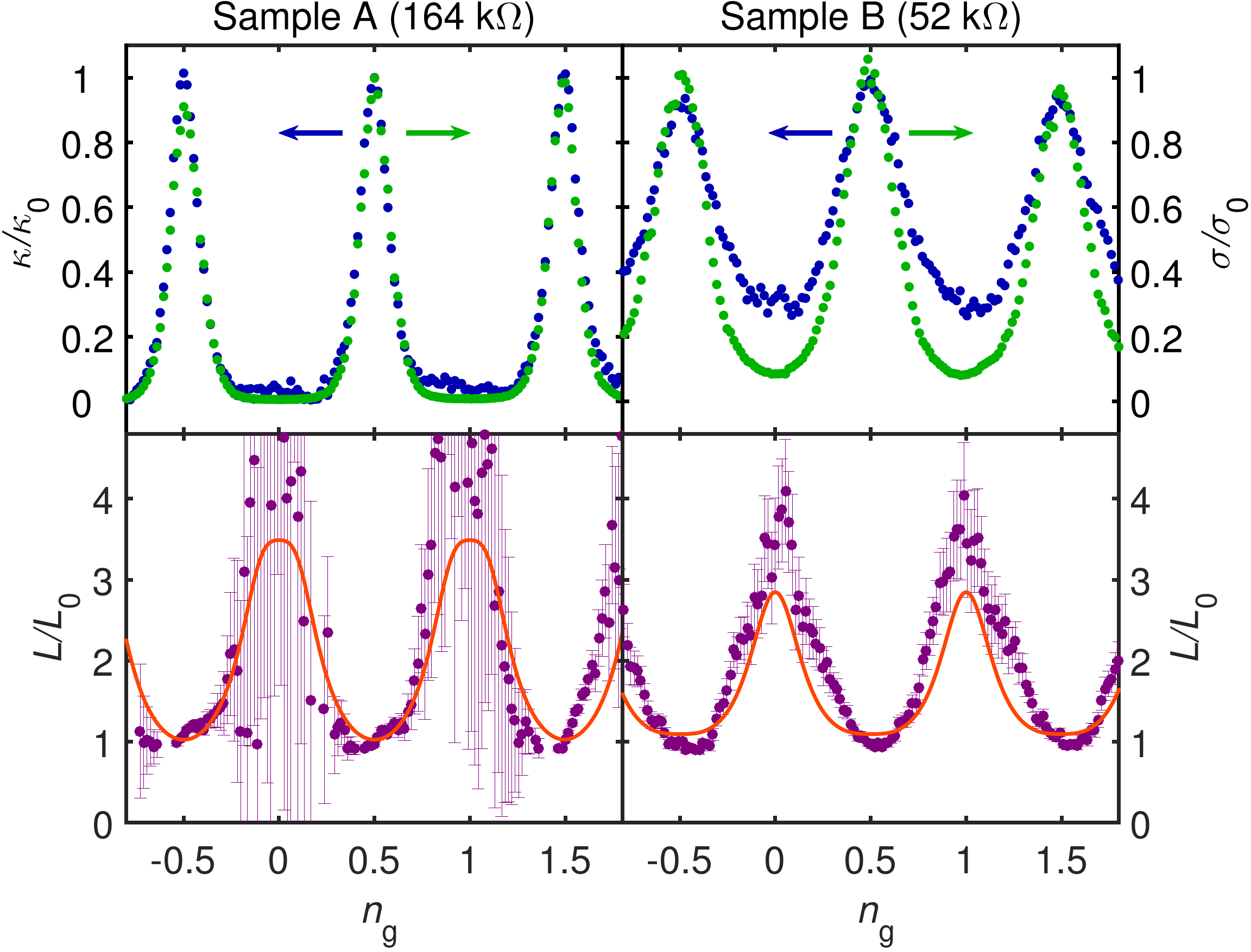}
\caption{Top: Thermal and electrical conductances of the SET for sample A (left) and sample B (right) at a bath temperature of 132 mK (sample A) and 152 mK (sample B). The thermal flow through the SET was calculated assuming that the Wiedemann-Franz law is fulfilled at gate-open state. The charge transport measurement was done at a bias of 22.4 $\mu$V (sample A) and 19.2 $\mu$V (sample B). The heat transport data was acquired by heating the source electronic bath by 60 mK (sample A) and 52 mK (sample B) above the bath temperature. Bottom: Lorenz ratio defined as $L/L_0$ where $L=\kappa/(\sigma \langle T_\text{m} \rangle)$ for sample A (left) and sample B (right). The red line is the theoretical prediction.} 
\label{fig:heatingregime}
\end{figure}

The analysis displayed in Fig.\ 3 of the main paper can be performed also in the heating regime, see Fig.\ S5. In that case, the temperature difference is large, preventing a direct analysis of the power through the SET in terms of thermal conductance. It is then particularly important to use in the calculation of the Lorenz factor, the temperature that is the average between the cold side (the bath temperature) and the hot side (the mean electron temperature). It shows a very good agreement with the behavior observed in main paper Fig.\ 3.

\begin{figure}[htb]
\includegraphics[width=0.5\columnwidth]{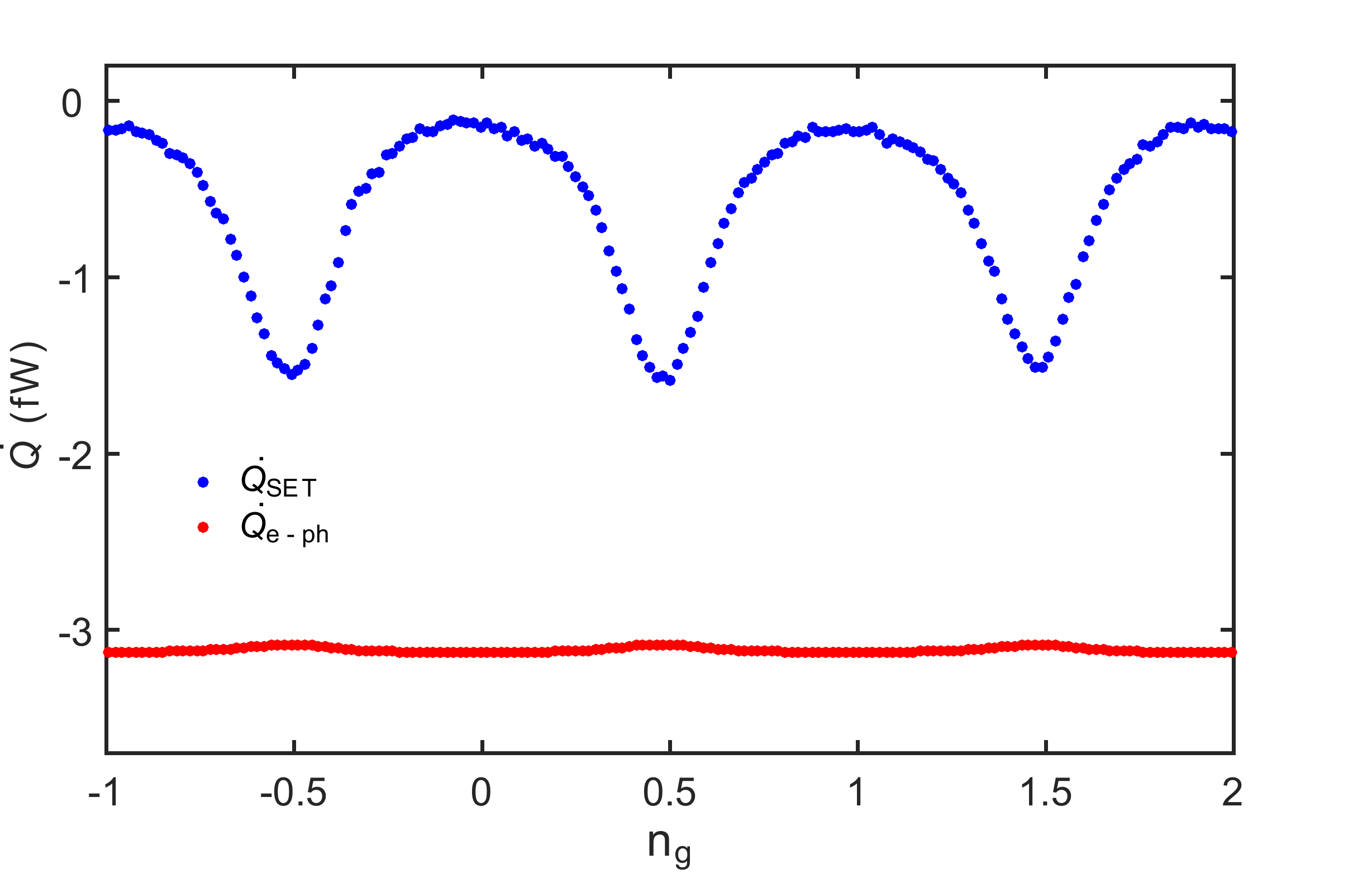}
\caption{Dependence of the electron-phonon coupling power $\dot{Q}_\text{e-ph}$ and the power flow through the SET $\dot{Q}_\text{SET}$, on the gate potential, related to the data of Fig.\ 2 in the main text.}
\label{fig:power}
\end{figure}

Figure S6 displays an example of the gate dependence of both the electron-phonon power and the power flowing through the SET. It is observed that the power flow through the SET can represent about 30$\%$ of the total power flow at 150 mK.

 \end{document}